\documentclass[aps,prc,groupedaddress,showpacs,showkeys]{revtex4}
\usepackage{graphicx}
\begin{document}
%
%
\title{Simple models for shell-model configuration densities}
%
%
\author{Edgar Ter\'an}
\author{Calvin W. Johnson}
\affiliation{San Diego State University}
%
%
\date{\today}
\begin{abstract}
We consider the secular behavior of shell-model configuration (partial) densities. 
When configuration densities are characterized by their moments, one often 
finds large third moments, which can make suitable parameterization of the
secular behavior problematic.  We review several parameterizations or models, 
and consider in depth three specific models: Cornish-Fisher, binomial, and 
modified Breit-Wigner distributions. Of these three the modified Breit-Wigner 
provides the best secular approximation to exact numerical configuration densities 
computed via full diagonalization from realistic interactions. 
\end{abstract}
%
\pacs{21zc}
%
\maketitle
%
\section{Introduction and Motivation}
%
Neutron-capture rates onto compound nuclear states are usually computed using
the statistical Hauser-Feshbach formalism\cite{HaFe52}. One important, and uncertain, input
into Hauser-Feshbach calculations is the density of excited states, or level density\cite{RaThKr97}; 
in addition to the total level density one also would 
like the exciton (particle-hole) density for
pre-equilibrium emission \cite{Gr66, StFaQa97, FeKeKo80,DeKo99}.

The nuclear level density is not trivial to extract experimentally, and there 
has recently been increasing theoretical efforts to compute and characterize 
the level density; it is beyond the scope of this paper to characterize all 
approaches, although some recent references are \cite{Pi94,
Go96,DeGo01,Or97,NaAl97,NaAl98,AlBeFa03,Hi04,HoGhZe04,NaFu05}.
Instead we focus on microscopic models, and in particular on the 
interacting shell model, which accurately describes \textit{low-lying} spectra 
and transitions for a broad range of nuclides. On
the other hand, to extract levels densities from traditional shell model
codes requires full diagonalization of the Hamiltonian, a computationally 
forbidding requirement. An alternative to diagonalization is the Monte
Carlo path integral technique \cite{Jo92}, which is well suited to 
thermal observables \cite{De95,Or97,NaAl97}. Although reasonable successful,
path integral methods are limited to interactions that are free of the 
``sign problem'' \cite{LaJoKoOr93,ADKLO94,KDL97}. Therefore we feel 
motivated to consider an alternate method based on spectral distribution 
theory (also known as statistical spectroscopy) \cite{MoFr75,Wo86}.

Spectral distribution theory computes the moments of the shell-model
Hamiltonian. One must then invert the moments to find the level density.
Note that most other methods to  compute the level density also 
require inversions, such as inverse Laplace transform 
through the saddle-point approximation\cite{NaAl97,NaAl98} or maximum entropy methods\cite{Or97}; 
in those approaches, as here, the success of the inversion depends 
upon the validity of explicit and implicit assumptions about the 
level density. For moment methods one must choose a parameterized 
``model'' for the secular behavior, and then adjust the parameters to 
fit the calculated moments; the choice of model implicitly makes assumptions
about the higher moments which were not fitted. 

One can model the total density of states as a sum of
partial or configuration densities \cite{KoPoSh86,HoKaZe03}. It is 
more efficient and tractable to compute the low moments of many 
configurations (subspaces), which one can do directly 
from the two-body matrix elements\cite{FrRa71,AyGi74,Wo86}, 
rather than many moments of the entire 
space. A further advantage is that one automatically gets out the 
particle-hole (exciton) density needed for calculating pre-equilibrium emission.

The success of a moments-method approach to the level density depends 
in large part on using a suitable model parameterization for the 
configuration density. 
Early attempts to find a suitable model for the level density utilized
Gram-Charlier and Edgeworth series representations in terms of derivatives 
of an asymptotic density \cite{ChZu72}; in 
essence one starts with a Gaussian\cite{MoFr75} and expands about it using 
orthogonal polynomials. Fixed-{\it J} expansions methods
have also been developed in terms of Gaussian shapes \cite{HaWo79,ZiBr81,AgKa97}.

Realistic configuration densities often have a large asymmetry, 
or third moment\cite{KoPoSh86,TeJo06}; an earlier study demonstrated that 
Gram-Charlier/Edgeworth expansions do poorly for large asymmetries\cite{KoPoSh86}.

In this paper we first describe the desired features of any 
distribution used to to model the configuration level density. 
We then review  several model parameterizations, focusing in particular 
on three: Cornish-Fisher, which had been shown superior to Gram-Charlier/Edgeworth
in \cite{KoPoSh86}; binomial distributions \cite{Zu01}; and finally a 
new proposal, a modified Breit-Wigner distribution. 
After comparing strengths and weaknesses of these three models, we 
illustrate their performance against exact shell-model 
calculations, with realistic interactions, of partial densities with 
a range of different asymmetries. We conclude that the modified Breit-Wigner (MBW) 
has significant advantages over the other two distributions. 
%
%
%
\section{\label{moments} Configuration moments}
Spectral distribution theory, or nuclear statistical spectroscopy,
analyzes many nuclear properties through low-lying moments of the 
Hamiltonian \cite{MoFr75,Wo86}. In this section we review the needed 
definitions; a somewhat more detail discussion is found in \cite{TeJo06}.
 We work in a finite model space
${\cal M}$ wherein the number of protons and neutrons is fixed. If in ${\cal
M}$ we represent the Hamiltonian as a matrix $\mathbf{H}$, then all
the moments can be written in terms of traces.
The total dimension of the space is $D = \mathrm{tr\,}\mathbf{1}$, and the
average is
$\langle \mathbf{O} \rangle = D^{-1} \mathrm{tr \,} \mathbf{O}.$
The first moment, or centroid, of the Hamiltonian is
$\bar{E} = \langle \mathbf{H} \rangle;$
all other moments are \textit{central} moments, computed relative to
the centroid:
\begin{equation}
\label{eq:shell_moms}
\mu^{(n)} = \langle (\mathbf{H}-\bar{E})^n \rangle \; \; ,\; \; n > 2 \;  \; . 
\end{equation}

The width $\sigma=\sqrt{\mu^{(2)}}$, and one scales
the higher moments by the width:
\begin{equation}
m^{(n)} = \frac{\mu^{(n)}}{\sigma^n}.
\end{equation}

In addition to the centroid and the width, the next two moments have
special names. The scaled third moment $m^{(3)}$ is the
\textit{asymmetry}, or the skewness; similarly, $m^{(4)}$-3 is the
\textit{excess} (hence a Gaussian has zero excess). 

We use $\alpha,
\beta, \gamma, \ldots$ to label subspaces. Let
\begin{equation}
P_\alpha =\sum_{i \in \alpha} \left | i \right \rangle \left \langle
i \right |
\end{equation}
be the projection operator for the $\alpha$-th subspace. In this paper 
we use spherical shell-model configurations to partition into 
subspaces; a configuration is all states of the form, e.g. 
$(1s_{1/2})_\pi^2 (1s_{1/2})_\nu^1 (0d_{3/2})_\pi^2 (0d_{3/2})_\nu^3$, 
etc.. One could use other group-theoretical partitions but spherical 
configurations have been the most widely studied for moment methods. 

Now one can introduce \textit{partial} or \textit{configuration densities},
\begin{equation}
\rho_\alpha(E) = \mathrm{tr} \, P_\alpha \delta(E-\mathbf{H} ).
\end{equation}
The  total density $\rho(E)$ is just the sum of the
partial densities. (Incidentally, one can differentiate between \textit{state} 
densities, which includes all $2J+1$ degeneracies in $J_z$, and 
\textit{level} densities, which do not. Be aware, however, that 
both terms are sometimes used interchangeably. While our discussion here is 
germane to both state and level densities, all our specific numerical 
examples refer to state densities.)

With projection operators for subspaces $P_\alpha$ in hand, we 
define partial or \textit{configuration moments}: the configuration dimension is
$D_\alpha = {\rm tr \,} P_\alpha$, the configuration centroid is
$\bar{E}_\alpha = D_\alpha^{-1} {\rm tr \,} P_\alpha \mathbf{H},$
while the configuration width $\sigma_\alpha$ and configuration
asymmetry $m_\alpha^{(3)}$ are defined in the obvious ways.
An early study of typical configuration moments 
with realistic interactions was done 
in \cite{KoPoSh86}, while more recently we conducted a detailed 
investigation\cite{TeJo06}.

One can compute the configuration moments directly from 
the many-body Hamiltonian, but that is computationally unfeasible 
for large systems. (For our examples herein, however, we do exactly that, 
generating the many-body Hamiltonian matrix via the shell-model 
code REDSTICK\cite{OrPr}.)  Alternately, one can make 
use of the available `analytic' formulae for configuration moments
\cite{FrRa71,AyGi74,Wo86}; these still require nontrivial computational 
effort, especially for third and fourth moments, but because they do 
not require the intermediate step of computing the many-body matrix 
elements, for large systems they are still faster. 

Finally, if one has the a secular  density $\rho(E)$ (which may model either a total density or a 
configuration density), one can compute the moments through integrals 
rather than traces: 
\begin{eqnarray}
D & = & \int dE \, \rho(E),  \\
\bar{E} &=& \frac{1}{D}\int dE \, E \rho(E), \\
\mu^{(2)} &=&  \frac{1}{D}\int dE \, (E-\bar{E})^2 \rho(E)
,  
\end{eqnarray}
and so on. Formally these integrals are equivalent to the trace definitions. 
\textit{An important question in this paper is how accurately the model 
parameterizations actually reproduce the desired moments,}  which we will address 
through numerical integration.

\section{\label{sec:models} Models for secular behavior}

We want a parameterized model $\rho_\mathrm{model}(E)$ for the secular behavior of 
the (configuration) density of states, with the parameters fixed by 
the low-lying moments of the density. That is, one finds the exact low-lying 
many-body configuration moments, the \textit{target moments}, from the Hamiltonian
and then finds the parameters of the secular model that, in principle, will reproduce those 
target moments. 
%
%
%
The \textit{ideal} characteristics for secular behavior are 

1.  Non-negative densities. Negative-valued densities are
of course unphysical. 

2.  Ease of deriving model parameters from target moments.  This is important 
when one has thousands  or even hundreds of thousands of 
configuration densities to construct. Furthermore, as discussed below, 
some `analytic' expressions for moments of model functions are in fact 
not accurate. 

3.  Fixed start- and end-points. While not essential, finite endpoints 
are useful and reflect the finite range of densities in a finite model space.

Not all models will contain all of these desirable characteristics. We
will discuss a total of five models. The first two models, 
Gram-Charlier/Edgeworth and exponential with polynomial 
argument (EPA), we discuss only briefly. The final three--
Cornish-Fisher, binomial, and modified Breit-Wigner--we 
consider in detail. 

Several of these models begin with a Gaussian with centroid $\bar{E}$ 
and width $\sigma$:
\begin{equation}
\rho(E) = \frac{1}{\sqrt{2\pi \sigma^2}} \exp \left( - \frac{1}{2} \left[ \frac{E-\bar{E}}{\sigma}\right]^2 \right).
\end{equation}
The Gram-Charlier/Edgeworth distributions \cite{Cr46} modify a Gaussian by 
multiplying it by a linear superposition of orthogonal Hermite polynomials. 
This intuitive approach has the highly desireable feature that, 
given the target 
moments, the parameters are trivial to determine. Unfortunately,
these distributions can also have unphysical, 
non-negative densities, which are particularly severe for highly asymmetric 
distributions. In addition, a previous study\cite{KoPoSh86} showed such 
distributions simply do poorly in reproducing the secular behavior of 
realistic densities. 

Another model which generalizes a Gaussian is the exponential with polynomial argument 
(EPA)  distribution\cite{GrMa95}, which takes the form 
$\rho(E) \sim \exp(aE + bE^2 +c E^3 + dE^4 )$. This 
distribution is positive definite, but has the problematic feature 
that the moments must be computed numerically. (In practice one uses a 
look-up table and interpolates\cite{EPA}).

For this paper we focus on three models:

1. Cornish-Fisher.  We use the representation found in Ref. \cite{KoPoSh86}, 
where it was studied and found to be superior to Gram-Charlier/Edgeworth 
distributions. We write it explicitly to show its connection to a Gaussian 
distribution:
\begin{eqnarray}
\rho(E) = \frac{1}{\sqrt{(2\pi)}} \exp \left( - \frac{1}{2} 
\left [a + b \frac{E-\bar{E}}{\sigma} + c
 \left ( \frac{E-\bar{E}}{\sigma} \right)^2 
+ d  \left ( \frac{E-\bar{E}}{\sigma} \right)^3 
\right ]^2 \right ) \nonumber \\
\times \left | \alpha + \beta 
 \frac{E-\bar{E}}{\sigma} 
+ \gamma 
\left ( \frac{E-\bar{E}}{\sigma} \right)^2 \right |,
\label{cfdist}
\end{eqnarray}
with the parameters
\begin{eqnarray}
a \approx \frac{m^{(4)}-3}{6}, \nonumber \\
b \approx    1 - \frac{1}{8} 
\left(m^{(4)}-3\right) - \frac{7}{36} \left(m^{(3)}\right)^2             , \nonumber \\
c \approx         -\frac{m^{(3)}}{3}            , \nonumber \\
d \approx    \frac{ \left( m^{(3)}\right)^2}{9} + \frac{m^{(4)}-3}{24}               , \nonumber \\
\alpha \approx   1 - \frac{7 \left(m^{(3)}\right)^2}{36} + \frac{m^{(4)}-3}{8}           , \nonumber \\
\beta \approx    -\frac{m^{(3)}}{3}           , \nonumber \\
\gamma \approx    \frac{\left( m^{(3)}\right)^2}{3} - \frac{m^{(4)}-3}{8}       
\label{cfparams}     .
\end{eqnarray}
While the Cornish-Fisher distribution is positive definite, and was shown \cite{KoPoSh86} 
to be superior to Gram-Charlier/Edgeworth distributions, it has some significant drawbacks. 
Most important is that the relations  (\ref{cfparams}) between the Cornish-Fisher 
parameters and the target moments are \textit{approximate}, derived assuming 
small deviations from a Gaussian.

\begin{figure}
\includegraphics[scale=0.43]{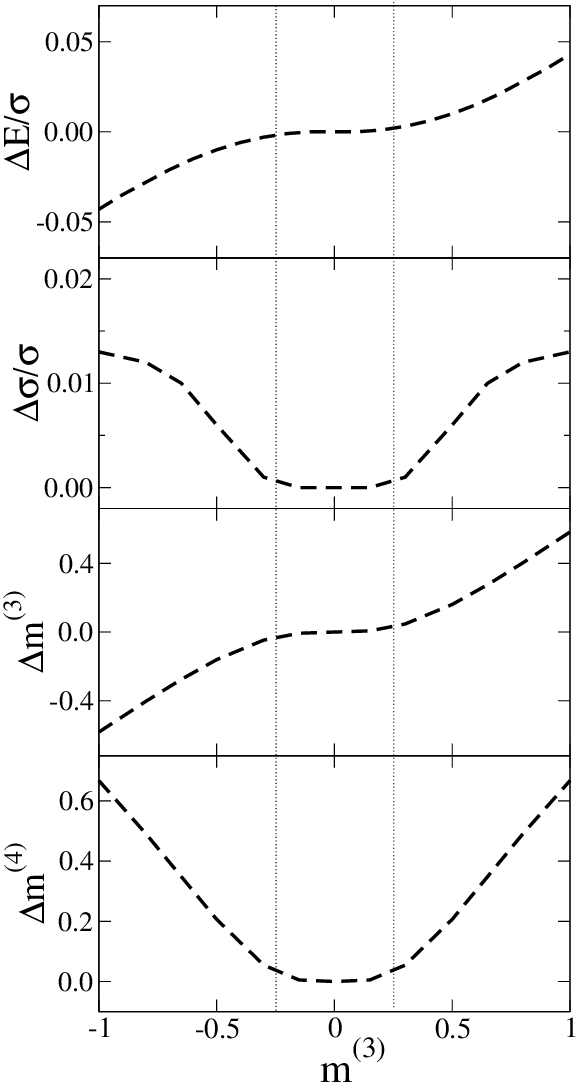}
\includegraphics[scale=0.43]{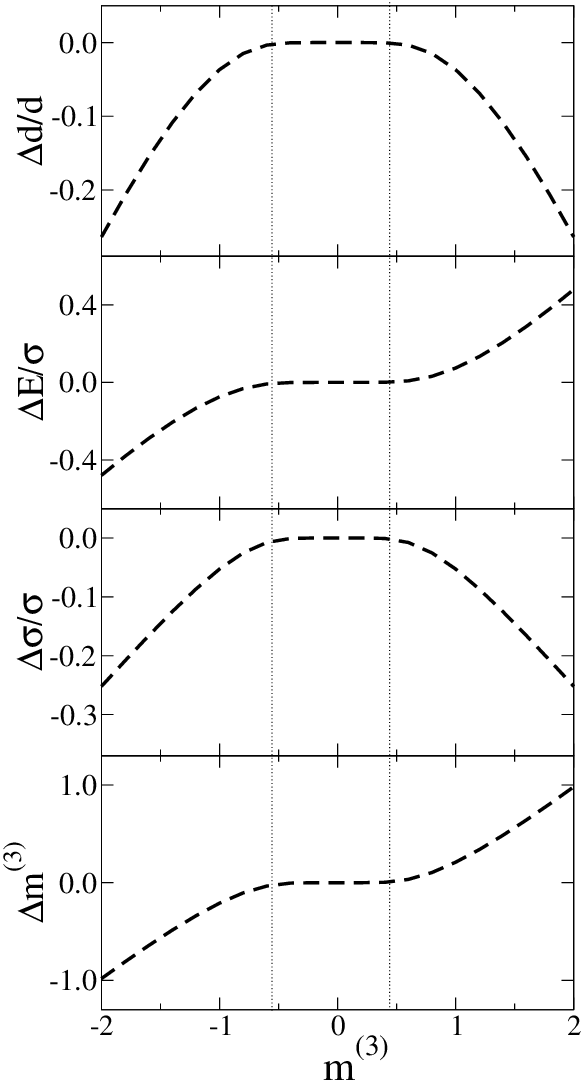}
\caption{\label{figure1} Moment discepancies in Cornish-Fisher (left) and continuous binomial (right)
distributions. Each graph corresponds to the numerical moments 
compared with target moments, at different asymmetries 
$m^{(3)}$. The plots for Cornish-Fisher are with $m^{(4)} = 3.0$ 
only;  distributions with higher fourth moments show even larger discrepancies.}
\end{figure}

We checked the consistency of the parameterization (\ref{cfparams}) by 
computing moments numerically. That is, given some target moments we used 
relations (\ref{cfparams}) to obtain the Cornish-Fisher parameters, and then 
obtained the numerical moments using (\ref{cfdist}). 
The lefthand side of Fig. \ref{figure1} shows 
the discrepancy between target and numerical moments. Not only do significant 
deviations develop beyond $| m^{(3)} > 0.5|$, the Cornish-Fisher distribution 
starts to develop unphysical bumps at the tails. 

2.  Binomial. The binomial distribution \cite{Zu01}
starts from the discrete expansion 
$(1+\lambda)^N = \sum{\lambda^k \left (
\begin{array}{c}
N \\ k
\end{array} \right )}$. If the excitation energies 
come in discrete steps $\epsilon$, that is, $E_x = k \epsilon$, one can compute the 
moments for the discrete distribution exactly:
\begin{eqnarray}
\bar{E} = \frac{N \epsilon \lambda}{1 + \lambda}  \nonumber \\
\mu^{(2)} = \frac{N \epsilon^2 \lambda}{(1 + \lambda)^2} \nonumber \\
m^{(3)} = \frac{1 - \lambda}{\sqrt{N \lambda}} \nonumber \\
m^{(4)} - 3 =\left( m^{(3)}\right )^2 - \frac{2}{N } \label{binomparams} \; .
\end{eqnarray}
One can obtain a continuous distribution by using gamma functions: 
\begin{equation}
\rho(E) = \lambda^{-E/\epsilon}
\frac{\Gamma(E_{max}/\epsilon+1)}
{\Gamma(E/\epsilon+1)\Gamma((E_{max}-E_x)/\epsilon+1) } \; ,
\end{equation}
where $E_\mathrm{max} = N\epsilon$. The distribution can be shifted 
by adding a simple displacement $E_\mathrm{min}$ to the energy. 
The parameters $N, \lambda$ are fitted to the third moment and to 
either the total dimension $d = (1+\lambda)^N$ or, through rescaling, 
to the fourth moment.

In the limit of no asymmetry and $N \rightarrow \infty$ the binomial 
distribution goes to a Gaussian. 
The binomial distribution is positive-definite and has finite 
endpoints at ($E_\mathrm{min},N\epsilon + E_\mathrm{min}$).  
Unfortunately, once again, the relation between the parameters and the 
target moments, exact for the discrete distribution, is only approximate 
for the \textit{continuous distribution}. We compared 
the discrepancy between target moments and numerical moment 
in the righthand side of Fig.~\ref{figure1}. Once again, the approximate 
formulas (\ref{binomparams}) is 
restricted to $|m^{(3)}| < 0.5$.

3. Modified Breit-Wigner (MBW). Given the problems with the 
previous models, we abandon generalizations of Gaussians and
instead propose
\begin{equation}
\rho(E) =\frac{1}{W^3} \frac{(E-E_{min})^2(E_{max}-E)^2}{(E-E_0)^2 + W^2}
\end{equation}
with endpoints given by $E_\mathrm{min},E_\mathrm{max}$.  
The first four moments can be computed analytically in terms 
of the four parameters; given target moments one can find 
the parameters by solving four nonlinear algebraic equations 
(plus an overall scale to match the correct dimension); 
we give the 
expressions in closed (but nontrivial) form in the Appendix. Unlike 
the Cornish-Fisher or binomial distributions, there is no 
discrepancy between the target and numerical moments. 

The MBW distribution is applicable for 
\begin{equation}
m^{(4)} -3 >1.42 \left(m^{(3)}\right)^2 - 0.52 \; .
\end{equation}
For third and fourth moments
inside the allowed area, one can find real solutions 
for the MBW parameters. 
 This region includes 
the numerically observed range for realistic interactions and model 
spaces \cite{TeJo06}. 

\begin{table}[t]
\caption{\label{table:moments-comparison}
Comparison of the moments reproduced by modified Breit-Wigner, Cornish-Fisher (C-F) 
and continuous binomial distributions. They are compared against exact calculations, 
all with GXPF1 interaction \cite{HoOtBrMi04} on {\it pf}-shell 
$^{44}$Ti for the same three configurations shown in Fig. \ref{figure2}.}
\begin{ruledtabular}
\begin{tabular}{ l   c c c c}
   &   & target &     C-F  &  Binomial    \\
\hline
Part 100 &
\begin{tabular}{ | l |}
 E             \\
 $\sigma$      \\
 $m^{(3)}$   \\
 $m^{(4)}$  \\
\end{tabular} &
\begin{tabular}{ c }
 -38.71  \\
  2.91   \\
  1.02   \\
  6.58   \\
\end{tabular} &
\begin{tabular}{ c }
  -38.76  \\
   2.87   \\
   1.26   \\
   6.87   \\
\end{tabular} &
\begin{tabular}{ c }
  -38.08   \\
   3.08    \\
   0.79    \\
   3.72    \\
\end{tabular} 
\\
\hline
Part 079 &
\begin{tabular}{ | l | }
 E          \\
 $\sigma$   \\
 $m^{(3)}$   \\
 $m^{(4)}$  \\
\end{tabular} &
\begin{tabular}{ c }
 -22.23    \\
  3.12     \\
 -0.70     \\
  5.43     \\
\end{tabular} &
\begin{tabular}{ c }
  -22.20  \\
   3.07   \\
  -0.61   \\
   3.59   \\
\end{tabular} &
\begin{tabular}{ c }
  -22.17   \\
   3.17    \\
  -0.62    \\
   3.36    \\
\end{tabular} \\ 
\hline
Part 019 &
\begin{tabular}{ | l | }
 E          \\
 $\sigma$   \\
 $m^{(3)}$   \\
 $m^{(4)}$  \\
\end{tabular} &
\begin{tabular}{ c }
 -27.39   \\
  3.30    \\
 -0.12    \\
  4.56    \\
\end{tabular} &
\begin{tabular}{ c }
  -27.39  \\
   3.30   \\
  -0.13   \\
   3.20   \\
\end{tabular} &
\begin{tabular}{ c }
  -27.39   \\
   3.30    \\
  -0.12    \\
   2.91    \\
\end{tabular} \\
\end{tabular} 
\end{ruledtabular}
\end{table}
\subsection{Direct comparison of model distributions}

To facilitate comparison of the Cornish-Fisher, binomial, and 
MBW distributions, we make two side-by-side comparisons of 
these three models along with exact numerical configuration densities; 
specifically,  we computed several 
partial densities for $^{44}$Ti (2 protons and 2 neutrons 
in a $pf$-shell valence space) with the realistic interaction GXPF1 \cite{HoOtBrMi04},
and chose to use three different configurations with different 
asymmetries $m^{(3)}$.

As discussed above, the ``analytic'' moments of the Cornish-Fisher and 
continuous binomial distribution are only approximate. 
Table \ref{table:moments-comparison} compares ``exact'' or target 
moments (shell-model calculations with REDSTICK code \cite{OrPr}) and the
numerical moments for the Cornish-Fisher and continuous binomial 
distributions. (The MBW numerical moments, as previously discussed, agree with the 
target moments.) 
The numerical Cornish-Fisher moments do well for centroids and widths, less 
well for $m^{(3)}$, and poorly for $m^{(4)}$; the continuous binomial 
has even larger errors. 

One could in principle adjust the Cornish-Fisher or 
binomial parameters (with a different, or even an exact, parameterization) to get 
agreement between target and numerical moments, but this would have to be 
done numerically or through a look-up table. Furthermore, the binomial 
distribution has a much smaller region of applicability than the MBW distribution;
one cannot find solutions for $|m^{(3)}|> 0.5$. 

\begin{figure}
\includegraphics[scale=0.73]{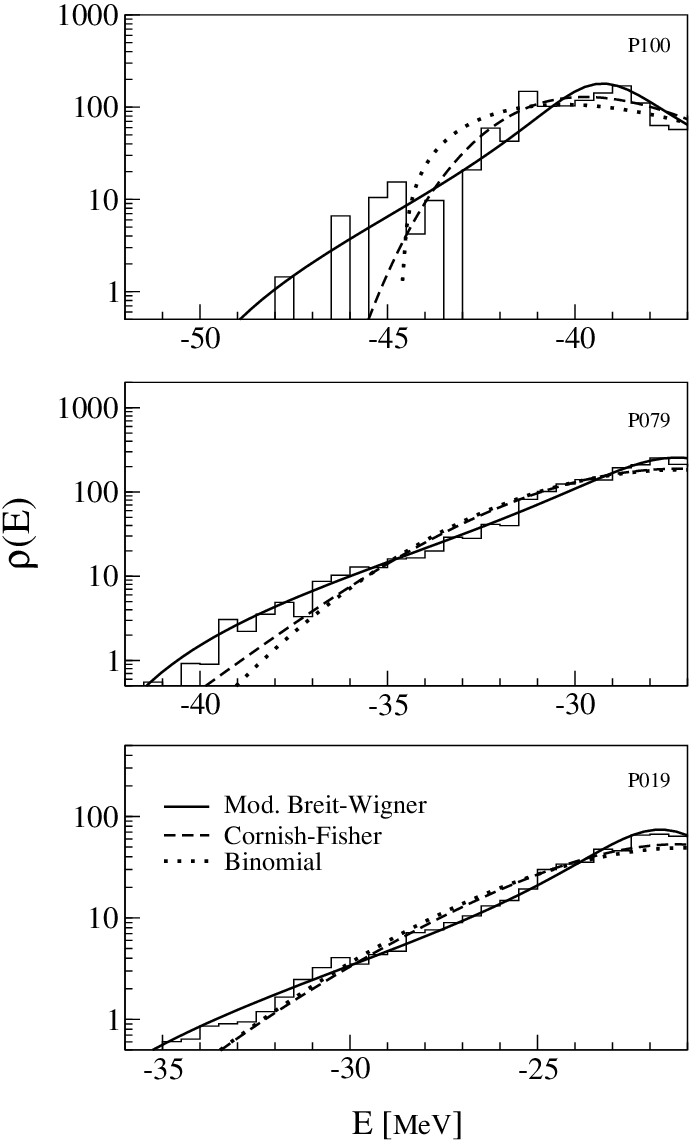}
\caption{\label{figure2} Representation of partial densities for three
different configurations of $^{44}$Ti in the pf-shell, with GXPF1 interaction.
The modified Breit-Wigner, Cornish-Fisher and Binomial distributions are compared
against the exact densities (solid plots with binned lines).}
\end{figure}

\begin{figure}
\includegraphics[scale=0.55]{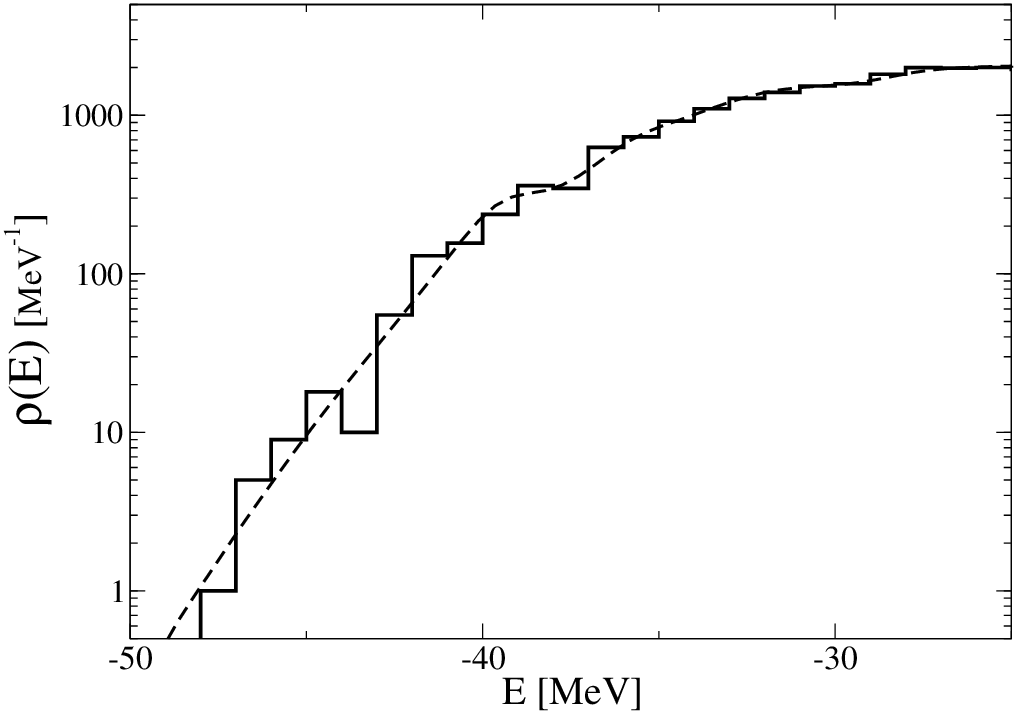}
\caption{\label{figure3} Full $^{44}$Ti level density in the {\it pf}-shell. 
The modeled level density (sum of MBW partial densities, broken line) is 
compared against the exact calculation (binned line).}
\end{figure}

In Fig.~\ref{figure2} we compare the MBW, Cornish-Fisher, and binomial 
distributions  against realistic configuration densities from 
direct diagonalization using REDSTICK.  Of the three models, 
the MBW appears to be better. 
Finally, we are able to compute the total level density with the MBW model
for $^{44}$Ti in the {\it pf}-shell with the GXPF1 interaction. Fig. 
\ref{figure3} shows such calculation, and the exact calculation with the
REDSTICK code, as well. In this example, the total level density is a sum
of a 100 partial densities.
\section{Summary}
In order to invert configuration moments to get level densities, 
one must posit a model distribution whose parameters are fixed 
by the low-lying moments. An important consideration for modeling 
partial or configuration densities are model distributions that 
perform well for large asymmetries (third moments) which we have 
previously shown to be important\cite{TeJo06}.
Continuing previous investigation\cite{KoPoSh86}, 
we have compared several distributions, in particular the Cornish-Fisher, 
continuous binomial, and a modified Breit-Wigner (MBW).  We suggest 
that the MBW, or perhaps some variant, may be the most reliable for configuration 
densities, especially at large asymmetries. 
\section{Acknowledgements}
This work is supported by grant DE-FG52-03NA00082  from the
Department  of Energy.
%
%
\appendix*
\section{Moments expressions for the modified Breit-Wigner distribution}

The modified Breit-Wigner distribution, with dimensions of [Energy]$^{-1}$, is
\begin{equation}
\rho(E) = \frac{1}{W^3}\frac{(E-E_{min})^2(E_{max}-E)^2}{(E-E_0)^2 + W^2}, 
\end{equation}
defined on the interval $E_\mathrm{min} \leq E \leq E_\mathrm{max}$.

\begin{widetext}

Define the dimensionless integrals
\begin{eqnarray}
I_0(a,b) = \int_a^b \frac{1}{x^2 +1} dx = \arctan b - \arctan a, \\
I_1(a,b) = \int_a^b \frac{x}{x^2 +1} dx = \frac{1}{2} \ln \left( \frac{b^2 +1}{a^2+1} \right ).
I_n(a,b) = \int_a^b \frac{x^n}{x^2 +1} dx .
\end{eqnarray}
Also define the general integral (with an easily proved recursion relation),
\begin{equation}
I_{n} = \int_a^b \frac{x^n}{x^2 +1} dx = \frac{1}{n-1}\left( b^{n-1} - a^{n-1} \right ) - I_{n-2},
\end{equation}
so that 
\begin{eqnarray} 
I_{2n} = (-1)^n I_0 + \sum_{k=1}^n \frac{(-1)^{n+k}}{2k-1} \left ( b^{2k-1} - a^{2k-1} \right ),\\
I_{2n+1} = (-1)^n I_1 + \sum_{k=1}^n \frac{(-1)^{n+k}}{2k} \left ( b^{2k} - a^{2k} \right ).
\end{eqnarray}

Now introduce the convenient dimensionful integral
\begin{equation}
M_n = \int_a^b \left(E-E_0 \right)^n \rho(E) dE.
\end{equation}
We can write this in terms of the dimensionless integrals, using 
$a = (E_\mathrm{min}-E_0)/W$ and $b = (E_\mathrm{max} - E_0)/W$, 
\begin{equation}
M_n = W^n \left(I_{n+4}-2(a+b)I_{n+3} +(a^2+4ab+b^2) I_{n+2}
-2(a+b)ab I_{n+1} + a^2b^2I_n \right).
\end{equation}
Finally, the moments in terms of the above expressions are
\begin{eqnarray}
\text{dimension    } D &=& \int_{a}^{b} \rho(E) dE = M_0 \\ 
\text{centroid     } \bar{E} &=& \int_{a}^{b} E \rho(E) dE = E_0 + \frac{M_1}{M_0} = E_0 + \Delta E \\ 
\text{variance     }{\sigma}^2 &=& \int_{a}^{b} (E-\bar{E})^2 \rho(E) dE = \frac{1}{M_0} 
         \left( M_2 - 2\Delta{E} M_1 + \Delta {E}^2 M_0 \right)   \\ 
  m^{(3)} &=& \int_{a}^{b} (E-\bar{E})^3 \rho(E) dE = \frac{1}{M_0 {\sigma}^3} 
        \left( M_3 - 3 \Delta {E} M_2 + 3 \Delta{E}^2 M_1 - \Delta{E}^3 M_0 \right)    \\ 
 m^{(4)} &=& \int_{a}^{b} (E-\bar{E})^4 \rho(E) dE = \frac{1}{M_0 {\sigma}^4} 
        \left( M_4 - 4 \Delta{E} M_3 + 6 \Delta{E}^2 M_2 - 4 \Delta{E}^3 M_1 + \Delta{E}^4 M_0 \right).    
\end{eqnarray}

\end{widetext}
%
%
\bibliography{biblio}

\begin{thebibliography}{38}
\expandafter\ifx\csname natexlab\endcsname\relax\def\natexlab#1{#1}\fi
\expandafter\ifx\csname bibnamefont\endcsname\relax
  \def\bibnamefont#1{#1}\fi
\expandafter\ifx\csname bibfnamefont\endcsname\relax
  \def\bibfnamefont#1{#1}\fi
\expandafter\ifx\csname citenamefont\endcsname\relax
  \def\citenamefont#1{#1}\fi
\expandafter\ifx\csname url\endcsname\relax
  \def\url#1{\texttt{#1}}\fi
\expandafter\ifx\csname urlprefix\endcsname\relax\def\urlprefix{URL }\fi
\providecommand{\bibinfo}[2]{#2}
\providecommand{\eprint}[2][]{\url{#2}}

\bibitem[{\citenamefont{Hauser and Feshbach}(1952)}]{HaFe52}
\bibinfo{author}{\bibfnamefont{W.}~\bibnamefont{Hauser}} \bibnamefont{and}
  \bibinfo{author}{\bibfnamefont{H.}~\bibnamefont{Feshbach}},
  \bibinfo{journal}{Phys. Rev.} \textbf{\bibinfo{volume}{87}},
  \bibinfo{pages}{366} (\bibinfo{year}{1952}).

\bibitem[{\citenamefont{{T. Rauscher, F.-K. Thielemann and K.-L.
  Kratz}}(1997)}]{RaThKr97}
\bibinfo{author}{\bibnamefont{{T. Rauscher, F.-K. Thielemann and K.-L.
  Kratz}}}, \bibinfo{journal}{Phys. Rev. C} \textbf{\bibinfo{volume}{56}},
  \bibinfo{pages}{1613} (\bibinfo{year}{1997}).

\bibitem[{\citenamefont{Griffin}(1966)}]{Gr66}
\bibinfo{author}{\bibfnamefont{J.~J.} \bibnamefont{Griffin}},
  \bibinfo{journal}{Phys. Rev. Lett.} \textbf{\bibinfo{volume}{17}},
  \bibinfo{pages}{478} (\bibinfo{year}{1966}).

\bibitem[{\citenamefont{{B. Strohmaier, M. Fassbender and S. M.
  Qaim}}(1997)}]{StFaQa97}
\bibinfo{author}{\bibnamefont{{B. Strohmaier, M. Fassbender and S. M. Qaim}}},
  \bibinfo{journal}{Phys. Rev. C} \textbf{\bibinfo{volume}{56}},
  \bibinfo{pages}{2654} (\bibinfo{year}{1997}).

\bibitem[{\citenamefont{{H. Feshbach, A. K. Kerman and S. E.
  Koonin}}(1980)}]{FeKeKo80}
\bibinfo{author}{\bibnamefont{{H. Feshbach, A. K. Kerman and S. E. Koonin}}},
  \bibinfo{journal}{Ann. Phys.} \textbf{\bibinfo{volume}{125}},
  \bibinfo{pages}{429} (\bibinfo{year}{1980}).

\bibitem[{\citenamefont{Dean and Koonin}(1999)}]{DeKo99}
\bibinfo{author}{\bibfnamefont{D.~J.} \bibnamefont{Dean}} \bibnamefont{and}
  \bibinfo{author}{\bibfnamefont{S.~E.} \bibnamefont{Koonin}},
  \bibinfo{journal}{Phys. Rev. C} \textbf{\bibinfo{volume}{60}},
  \bibinfo{pages}{054306} (\bibinfo{year}{1999}).

\bibitem[{\citenamefont{Ormand}(1997)}]{Or97}
\bibinfo{author}{\bibfnamefont{W.~E.} \bibnamefont{Ormand}},
  \bibinfo{journal}{Phys. Rev. C} \textbf{\bibinfo{volume}{56}},
  \bibinfo{pages}{R1678} (\bibinfo{year}{1997}).

\bibitem[{\citenamefont{{B. Pichon}}(1994)}]{Pi94}
\bibinfo{author}{\bibnamefont{{B. Pichon}}}, \bibinfo{journal}{Nucl. Phys. A}
  \textbf{\bibinfo{volume}{568}}, \bibinfo{pages}{553} (\bibinfo{year}{1994}).

\bibitem[{\citenamefont{Goriely}(1996)}]{Go96}
\bibinfo{author}{\bibfnamefont{S.}~\bibnamefont{Goriely}},
  \bibinfo{journal}{Nucl. Phys. A} \textbf{\bibinfo{volume}{605}},
  \bibinfo{pages}{28} (\bibinfo{year}{1996}).

\bibitem[{\citenamefont{{P. Demetriou and S. Goriely}}(2001)}]{DeGo01}
\bibinfo{author}{\bibnamefont{{P. Demetriou and S. Goriely}}},
  \bibinfo{journal}{Nucl. Phys. A} \textbf{\bibinfo{volume}{695}},
  \bibinfo{pages}{95} (\bibinfo{year}{2001}).

\bibitem[{\citenamefont{Nakada and Alhassid}(1998)}]{NaAl98}
\bibinfo{author}{\bibfnamefont{H.}~\bibnamefont{Nakada}} \bibnamefont{and}
  \bibinfo{author}{\bibfnamefont{Y.}~\bibnamefont{Alhassid}},
  \bibinfo{journal}{Phys. Lett. B} \textbf{\bibinfo{volume}{436}},
  \bibinfo{pages}{231} (\bibinfo{year}{1998}).

\bibitem[{\citenamefont{Nakada and Alhassid}(1997)}]{NaAl97}
\bibinfo{author}{\bibfnamefont{H.}~\bibnamefont{Nakada}} \bibnamefont{and}
  \bibinfo{author}{\bibfnamefont{Y.}~\bibnamefont{Alhassid}},
  \bibinfo{journal}{Phys. Rev. Lett.} \textbf{\bibinfo{volume}{79}},
  \bibinfo{pages}{2939} (\bibinfo{year}{1997}).

\bibitem[{\citenamefont{{S. Hilaire}}(2004)}]{Hi04}
\bibinfo{author}{\bibnamefont{{S. Hilaire}}}, \bibinfo{journal}{Phys. Lett. B}
  \textbf{\bibinfo{volume}{583}}, \bibinfo{pages}{264} (\bibinfo{year}{2004}).

\bibitem[{\citenamefont{{Y. Alhassid, G. F. Bertsch, and L.
  Fang}}(2003)}]{AlBeFa03}
\bibinfo{author}{\bibnamefont{{Y. Alhassid, G. F. Bertsch, and L. Fang}}},
  \bibinfo{journal}{Phys. Rev. C} \textbf{\bibinfo{volume}{68}},
  \bibinfo{pages}{044322} (\bibinfo{year}{2003}).

\bibitem[{\citenamefont{{M. Horoi, M. Ghita, and V.
  Zelevinsky}}(2004)}]{HoGhZe04}
\bibinfo{author}{\bibnamefont{{M. Horoi, M. Ghita, and V. Zelevinsky}}},
  \bibinfo{journal}{Phys. Rev. C} \textbf{\bibinfo{volume}{69}},
  \bibinfo{pages}{041307} (\bibinfo{year}{2004}).

\bibitem[{\citenamefont{{H. Nakamura and T. Fukahori}}(2005)}]{NaFu05}
\bibinfo{author}{\bibnamefont{{H. Nakamura and T. Fukahori}}},
  \bibinfo{journal}{Phys. Rev. C} \textbf{\bibinfo{volume}{72}},
  \bibinfo{pages}{064329} (\bibinfo{year}{2005}).

\bibitem[{\citenamefont{{C. W. Johnson, S. E. Koonin, G. H. Lang and W. E.
  Ormand}}(1995)}]{Jo92}
\bibinfo{author}{\bibnamefont{{C. W. Johnson, S. E. Koonin, G. H. Lang and W.
  E. Ormand}}}, \bibinfo{journal}{Phys. Rev. Lett.}
  \textbf{\bibinfo{volume}{69}}, \bibinfo{pages}{3157} (\bibinfo{year}{1995}).

\bibitem[{\citenamefont{{D. J. Dean, S. E. Koonin, K. Langanke, P. B. Radha and
  Y. Alhassid}}(1995)}]{De95}
\bibinfo{author}{\bibnamefont{{D. J. Dean, S. E. Koonin, K. Langanke, P. B.
  Radha and Y. Alhassid}}}, \bibinfo{journal}{Phys. Rev. Lett.}
  \textbf{\bibinfo{volume}{74}}, \bibinfo{pages}{2909} (\bibinfo{year}{1995}).

\bibitem[{\citenamefont{{ G. H. Lang, C. W. Johnson, S. E. Koonin, and W. E.
  Ormand}}(1993)}]{LaJoKoOr93}
\bibinfo{author}{\bibnamefont{{ G. H. Lang, C. W. Johnson, S. E. Koonin, and W.
  E. Ormand}}}, \bibinfo{journal}{Phys. Rev. C} \textbf{\bibinfo{volume}{48}},
  \bibinfo{pages}{1518} (\bibinfo{year}{1993}).

\bibitem[{\citenamefont{{Y. Alhassid, D. J. Dean, S. E. Koonin, G. Lang, and W.
  E. Ormand}}(1994)}]{ADKLO94}
\bibinfo{author}{\bibnamefont{{Y. Alhassid, D. J. Dean, S. E. Koonin, G. Lang,
  and W. E. Ormand}}}, \bibinfo{journal}{Phys. Rev. Lett.}
  \textbf{\bibinfo{volume}{72}}, \bibinfo{pages}{613} (\bibinfo{year}{1994}).

\bibitem[{\citenamefont{{S. E. Koonin, D. J. Dean, and K.
  Langanke}}(1997)}]{KDL97}
\bibinfo{author}{\bibnamefont{{S. E. Koonin, D. J. Dean, and K. Langanke}}},
  \bibinfo{journal}{Phys. Rep.} \textbf{\bibinfo{volume}{278}},
  \bibinfo{pages}{2} (\bibinfo{year}{1997}).

\bibitem[{\citenamefont{{K. K. Mon and J. B. French}}(1975)}]{MoFr75}
\bibinfo{author}{\bibnamefont{{K. K. Mon and J. B. French}}},
  \bibinfo{journal}{Ann. Phys.} \textbf{\bibinfo{volume}{95}},
  \bibinfo{pages}{90} (\bibinfo{year}{1975}).

\bibitem[{\citenamefont{Wong}(1986)}]{Wo86}
\bibinfo{editor}{\bibfnamefont{S.~S.~M.} \bibnamefont{Wong}}, ed.,
  \emph{\bibinfo{title}{Nuclear Statistical Spectroscopy}}
  (\bibinfo{publisher}{Oxford University Press}, \bibinfo{address}{New York},
  \bibinfo{year}{1986}).

\bibitem[{\citenamefont{{V. K. B. Kota, V. Potbhare and P.
  Shenoy}}(1986)}]{KoPoSh86}
\bibinfo{author}{\bibnamefont{{V. K. B. Kota, V. Potbhare and P. Shenoy}}},
  \bibinfo{journal}{Phys. Rev. C} \textbf{\bibinfo{volume}{34}},
  \bibinfo{pages}{2330} (\bibinfo{year}{1986}).

\bibitem[{\citenamefont{Horoi et~al.}(2003)\citenamefont{Horoi, Kaiser, and
  Zelevinsky}}]{HoKaZe03}
\bibinfo{author}{\bibfnamefont{M.}~\bibnamefont{Horoi}},
  \bibinfo{author}{\bibfnamefont{J.}~\bibnamefont{Kaiser}}, \bibnamefont{and}
  \bibinfo{author}{\bibfnamefont{V.}~\bibnamefont{Zelevinsky}},
  \bibinfo{journal}{Phys. Rev. C} \textbf{\bibinfo{volume}{67}},
  \bibinfo{pages}{054309} (\bibinfo{year}{2003}).

\bibitem[{\citenamefont{{J. B. French and K. F. Ratcliff}}(1971)}]{FrRa71}
\bibinfo{author}{\bibnamefont{{J. B. French and K. F. Ratcliff}}},
  \bibinfo{journal}{Phys. Rev. C} \textbf{\bibinfo{volume}{3}},
  \bibinfo{pages}{94} (\bibinfo{year}{1971}).

\bibitem[{\citenamefont{{S. Ayik and J.N. Ginocchio}}(1974)}]{AyGi74}
\bibinfo{author}{\bibnamefont{{S. Ayik and J.N. Ginocchio}}},
  \bibinfo{journal}{Nucl. Phys. A} \textbf{\bibinfo{volume}{221}},
  \bibinfo{pages}{285} (\bibinfo{year}{1974}).

\bibitem[{\citenamefont{{F. S. Chang and A. Zuker}}(1972)}]{ChZu72}
\bibinfo{author}{\bibnamefont{{F. S. Chang and A. Zuker}}},
  \bibinfo{journal}{Nucl. Phys. A} p. \bibinfo{pages}{417}
  (\bibinfo{year}{1972}).

\bibitem[{\citenamefont{{R. U. Haq and S. S. M. Wong}}(1979)}]{HaWo79}
\bibinfo{author}{\bibnamefont{{R. U. Haq and S. S. M. Wong}}},
  \bibinfo{journal}{Nucl. Phys. A} \textbf{\bibinfo{volume}{327}},
  \bibinfo{pages}{314} (\bibinfo{year}{1979}).

\bibitem[{\citenamefont{{M. R. Zirnbauer and D. M. Brink}}(1981)}]{ZiBr81}
\bibinfo{author}{\bibnamefont{{M. R. Zirnbauer and D. M. Brink}}},
  \bibinfo{journal}{Z. Phys. A} \textbf{\bibinfo{volume}{301}},
  \bibinfo{pages}{237} (\bibinfo{year}{1981}).

\bibitem[{\citenamefont{Agrawal and Kataria}(1997)}]{AgKa97}
\bibinfo{author}{\bibfnamefont{B.~K.} \bibnamefont{Agrawal}} \bibnamefont{and}
  \bibinfo{author}{\bibfnamefont{S.~K.} \bibnamefont{Kataria}},
  \bibinfo{journal}{Z. Phys. A} \textbf{\bibinfo{volume}{356}},
  \bibinfo{pages}{369} (\bibinfo{year}{1997}).

\bibitem[{\citenamefont{Teran and Johnson}(2006)}]{TeJo06}
\bibinfo{author}{\bibfnamefont{E.}~\bibnamefont{Teran}} \bibnamefont{and}
  \bibinfo{author}{\bibfnamefont{C.~W.} \bibnamefont{Johnson}},
  \bibinfo{journal}{Phys. Rev. C} \textbf{\bibinfo{volume}{73}},
  \bibinfo{pages}{024303} (\bibinfo{year}{2006}).

\bibitem[{\citenamefont{Zuker}(2001)}]{Zu01}
\bibinfo{author}{\bibfnamefont{A.~P.} \bibnamefont{Zuker}},
  \bibinfo{journal}{Phys. Rev. C} \textbf{\bibinfo{volume}{64}},
  \bibinfo{pages}{021303(R)} (\bibinfo{year}{2001}).

\bibitem[{\citenamefont{Ormand}(2004)}]{OrPr}
\bibinfo{author}{\bibfnamefont{W.~E.} \bibnamefont{Ormand}}
  (\bibinfo{year}{2004}), \bibinfo{note}{private communication}.

\bibitem[{\citenamefont{{H. Cramer}}(1946)}]{Cr46}
\bibinfo{author}{\bibnamefont{{H. Cramer}}},
  \emph{\bibinfo{title}{{Mathematical Methods of Statistics}}}
  (\bibinfo{publisher}{{Princeton Univ. Press}},
  \bibinfo{address}{{Princeton}}, \bibinfo{year}{1946}).

\bibitem[{\citenamefont{{S. M. Grimes and T. N. Massey}}(1995)}]{GrMa95}
\bibinfo{author}{\bibnamefont{{S. M. Grimes and T. N. Massey}}},
  \bibinfo{journal}{Phys. Rev. C} \textbf{\bibinfo{volume}{51}},
  \bibinfo{pages}{606} (\bibinfo{year}{1995}).

\bibitem[{\citenamefont{{T. N. Massey}}()}]{EPA}
\bibinfo{author}{\bibnamefont{{T. N. Massey}}}, \bibinfo{note}{{private
  communication}}.

\bibitem[{\citenamefont{Honma et~al.}(2004)\citenamefont{Honma, Otsuka, Brown,
  and Mizusaki}}]{HoOtBrMi04}
\bibinfo{author}{\bibfnamefont{M.}~\bibnamefont{Honma}},
  \bibinfo{author}{\bibfnamefont{T.}~\bibnamefont{Otsuka}},
  \bibinfo{author}{\bibfnamefont{B.~A.} \bibnamefont{Brown}}, \bibnamefont{and}
  \bibinfo{author}{\bibfnamefont{T.}~\bibnamefont{Mizusaki}},
  \bibinfo{journal}{Physical Review C} \textbf{\bibinfo{volume}{69}},
  \bibinfo{eid}{034335} (\bibinfo{year}{2004}).

\end{thebibliography}
\end{document}